\documentclass[prb,twocolumn,showpacs,floatfix]{revtex4-1}
\usepackage{graphicx}
\usepackage{amssymb}
\usepackage{amsmath}
\usepackage{xspace}
\usepackage{url}

\begin{document} %%%%%%%%%%%%%%%%%%%%%%%%%%%%%%%%%%%%%%%%%%%%%%%%%%%%%%%%%%
%------------------------------------------------------------------------------ 
% Title
%------------------------------------------------------------------------------
\title{Physisorption of an electron in deep surface potentials off a dielectric surface}

%------------------------------------------------------------------------------ 
% Authors
%------------------------------------------------------------------------------
%------------------------------------------------------------------------------ 
% Date
%------------------------------------------------------------------------------
\author{R. L. Heinisch, F. X. Bronold, and 
H. Fehske}
\affiliation{Institut f{\"ur} Physik,
             Ernst-Moritz-Arndt-Universit{\"a}t Greifswald,
             17489 Greifswald,
             Germany}

\date{\today}
\begin{abstract}
We study phonon-mediated adsorption and desorption of an electron at dielectric surfaces 
with deep polarization-induced surface potentials where multi-phonon transitions 
are responsible for electron energy relaxation. Focusing on 
multi-phonon processes due to the nonlinearity of the coupling between the external electron 
and the acoustic bulk phonon triggering the transitions between surface states, we calculate 
electron desorption times for graphite, MgO, CaO, \(\text{Al}_2\text{O}_3\), and \(\text{SiO}_2\) 
and electron sticking coefficients for \(\text{Al}_2\text{O}_3\), CaO, and \(\text{SiO}_2\). To 
reveal the kinetic stages of electron physisorption, we moreover study the time evolution of the 
image state occupancy and the energy-resolved desorption flux. Depending on the potential depth and 
the surface temperature we identify two generic scenarios: (i) adsorption via trapping in shallow 
image states followed by relaxation to the lowest image state and desorption from that state 
via a cascade through the second strongly bound image state in not too deep potentials and (ii) 
adsorption via trapping in shallow image states but followed by a relaxation bottleneck retarding 
the transition to the lowest image state and desorption from that state via a one step process to 
the continuum in deep potentials.
\end{abstract}
%\pacs{52.40.Hf, 73.20.-r, 68.43.Mn}
\maketitle

\section{Introduction}

Image states, arising from the polarization-induced interaction between an electron and a surface, offer the possibility 
for electron trapping at a surface. Since their original prediction\cite{CC69} for the surface of liquid and solid He, 
they have been extensively studied for metallic surfaces.\cite{DAG84,SH84,Fauster94,HSR97,Hoefer99} But image states 
also exist for dielectric
surfaces provided the electron affinity of the dielectric is negative, that is, the vacuum level falls inside the gap 
between the valence and the conduction band. Image states are then the lowest unoccupied states and should hence allow for 
temporary trapping of external electrons. So far image states at a dielectric surface have been only observed for 
graphite,\cite{LMT99} but they are expected for other dielectrics with negative electron affinity as well, for instance, 
boron nitride\cite{LSG99} and the alkaline earth oxides.\cite{BKP07}

Based on the idea of a two-dimensional electron surface plasma,\cite{EC87,EC88,BBD97,GMB02} electron trapping in 
image states has been suspected for a long time to be responsible for the build-up of surface charges at plasma 
walls. We have recently proposed therefore to consider the charging of a plasma wall as an electron physisorption 
process.\cite{BFKD08,BDF09} Indeed, for plasma walls with negative electron affinity image states should contribute 
to the very beginning of the charging process, when the wall carries no charges yet and the image states thus fall  
inside the energy gap of the wall. Only with increasing surface charge image states are expected to play a less 
important role because the Coulomb barrier due to the electrons already residing on the wall shifts image states 
to an energy range where they are destabilized by unoccupied bulk states. The later stages of charge
collection most probably occurs via surface resonances or empty volume states.~\cite{BHMF10}

Regardless of its importance for charge collection at dielectric plasma walls, the electron kinetics in the 
image states of a dielectric surface is an interesting phenomenon in its own right. In addition, it is relevant 
in other physical contexts as well. For instance,
(i) in electron emitters, such as cesium-doped silicon oxide films with negative electron affinity, electron 
emission via image states reduces the operational voltage considerably,\cite{GDK05} (ii) in gallium arsenide based 
heterostructures surface charging can be used for the contactless gating of field devices,\cite{BGY05} and (iii) for 
the alkaline earth oxides, studied in the field of heterogeneous catalysis,\cite{Hattori95,Hattori03,WE00,Freund07} 
the electronic surface states provide the environment for catalytic reactions. Some situations encompass electronic 
transitions from bulk to surface states, as it is the case for electron emitters, while for others, the electron does 
not penetrate into the bulk and the electron kinetics takes only place in surface states. Interesting questions in 
this case are the probability for temporary trapping in these states, the mechanism of electron energy relaxation at 
the surface, and the time after which a trapped electron is released.

This is the concluding paper out of a series of three on the phonon-mediated physisorption of an electron in the 
image states of a dielectric surface. As in our previous work, Refs.~\onlinecite{HBF10a,HBF10b} (thereafter 
referred to as I and II), we investigate adsorption and desorption of an electron at finite temperatures 
assuming an acoustic longitudinal bulk phonon to control energy relaxation at the surface. 
For the dielectric material we are considering, the level spacing of the lowest two bound states typically exceeds 
the Debye energy, implying that multi-phonon processes have 
to be taken into account. In I and II, we have studied desorption and sticking using an expansion of the energy 
dependent T matrix,\cite{BY73,GKT80b,AM91} allowing to calculate one- and two-phonon 
transition probabilities. This approach is however limited to very few materials, for instance, graphite and MgO. In 
the following we will adopt a different strategy, calculating multi-phonon transition probabilities due to the nonlinear 
electron-phonon coupling non-perturbatively. This allows us to calculate the desorption time and the sticking 
coefficient for the deeper surface potentials of CaO, \(\text{Al}_2\text{O}_3\) and \(\text{SiO}_2\). 

The remaining paper is structured as follows. In Sec. \ref{Electron kinetics} we briefly recall the quantum-kinetic 
approach to physisorption. In Sec. \ref{Electron-surface interaction and transition rates}, we calculate the multi-phonon 
state-to-state transition probabilities. In Sec. \ref{Results} we present our results for the desorption time and the 
prompt and kinetic energy-resolved and energy-averaged sticking coefficient. In this section we also discuss the 
time evolution of the bound state occupancy and the energy-resolved desorption flux. Section 
\ref{Discussion: Two  state system} is devoted to the analytic treatment of a simplified two-state model, used to 
identify two generic physisorption scenarios into which we can classify the results of this paper as well as 
our previous results, before we conclude in Sec. \ref{Summary}.

\section{Electron kinetics}
\label{Electron kinetics}

As in I~\cite{HBF10a} and II~\cite{HBF10b} we describe the time evolution of the occupancy of the 
bound surface states with a quantum-kinetic rate equation.\cite{GKT80a,KG86} It captures all three characteristic 
stages of physisorption:~\cite{IN76,Brenig82} initial trapping, subsequent relaxation and desorption. 

The time dependence of the occupancies of the bound states is given by\cite{GKT80a,KG86} 
\begin{align}
\frac{\mathrm{d}}{\mathrm{d}t}n_n(t)=&\sum_{n^\prime} \left[W_{n n^\prime} n_{n^\prime}(t) - W_{n^\prime n} n_n(t) \right] \nonumber \\
& -\sum_k W_{k n} n_n(t) +\sum_k \tau_t W_{nk} j_k(t)  \label{fullrateeqn} \\
=& \sum_{n^\prime} T_{nn^\prime} n_{n^\prime}(t)+\sum_k \tau_t W_{nk} j_k(t) \text{ ,} \label{fullrateeqn2}
\end{align}
where \(W_{n^\prime n}\) is the probability per unit time for a transition from a bound state \(n\) to another bound 
state \(n^\prime\), \(W_{kn}\) and \(W_{nk}\) are the probabilities per unit time for a transition from the bound 
state \(n\) to the continuum state \(k\) and vice versa and \(\tau_t=2L/v_z\) is the transit time through the 
surface potential of width \(L\), which, in the limit \(L\rightarrow \infty\), can be absorbed into the transition 
probability. The matrix \(T_{nm}\) is defined implicitly by the above equation. The last term in 
Eqs. (\ref{fullrateeqn}) and (\ref{fullrateeqn2}), respectively, gives the increase in the bound state occupancy due 
to trapping of an electron in bound surface states. 

The probability for an approaching electron in the continuum state \(k\) to make a transition to any of the bound states 
is given by the prompt energy-resolved sticking coefficient,
\begin{align}
s_{e,k}^\text{prompt}=\tau_t\sum_n W_{nk} \text{.}
\end{align}

Treating the incident electron flux as an externally specified parameter, the solution of Eq. (\ref{fullrateeqn}) describes the subsequent relaxation and desorption. It is given by
\begin{align}
n_n(t)=\sum_{\kappa} e^{-\lambda_\kappa t} \int_{-\infty}^t \mathrm{d}t^\prime e^{\lambda_\kappa t^\prime} e_n^{(\kappa)} \sum_{kl} \tilde{e}_l^{(\kappa)} \tau_t W_{lk} j_k(t^\prime) \text{ ,} \label{rateqensolution}
\end{align}
where \(e_n^{(\kappa)} \) and \(\tilde{e}_n^{(\kappa)}\) are the right and left eigenvectors to the eigenvalue \(-\lambda_\kappa\) of the matrix \(\mathbf{T}\).

If the modulus of one eigenvalue, \(\lambda_0\), is considerably smaller than the moduli of the other eigenvalues, 
\(\lambda_\kappa\), a unique desorption time and a unique sticking coefficient can be identified.\cite{KG86} In 
this case \(\lambda_0\) governs the long time behavior of the equilibrium occupation of the bound states, 
\(n_q^\mathrm{eq} \sim e^{- E_q/k_BT_s}\), and its inverse can be identified with the desorption time, \(\lambda_0^{-1}=\tau_e\). In this case the bound state occupancy  \(n_n(t)\) splits into a slowly varying part \(n_n^0(t)\) given by the \(\kappa =0\) summand in Eq. (\ref{rateqensolution}) and a quickly varying part  \(n_n^f(t)\) given by the sum over \(\kappa \neq0\) in Eq. (\ref{rateqensolution}).

The "adsorbate``, i.e. the fraction of the trapped electron remaining in the surface states for times on the order of the desorption time, is given by the slowly varying part only, \(n^0(t)=\sum_n n_n^0(t)\). 
Differentiating \(n^0(t)\) with respect to the time, 
\begin{align}
\frac{d}{dt}{n}^0(t)=\sum_k s_ {e,k}^\text{kinetic} j_k(t) - \lambda_0 n^0(t) \text{ ,}
\end{align}
we can, following Brenig\cite{Brenig82}, identify the kinetic energy-resolved sticking coefficient
\begin{align}
s_{e,k}^\text{kinetic}=\tau_t \sum_{n,n^\prime}  e_{n^\prime}^{(0)} \tilde{e}_n^{(0)} W_{n k} \text{ ,}
\end{align}
giving the probability for both, initial trapping and subsequent relaxation. 

If the incident unit electron flux corresponds to an electron with Boltzmann distributed kinetic energies, the prompt or kinetic energy-averaged sticking coefficient is given by
\begin{align}
s_e^{\dots}=\frac{\sum_k s_{e,k}^{\dots} k e^{-\beta_e E_k}}{\sum_k k e^{-\beta_e E_k}} \text{ ,}
\end{align}
where \(\beta_e^{-1}=k_B T_e\) is the mean electron energy.

The desorption flux, that is, the flux due to an electron that is not instantly reflected at the boundary but sticks to the surface for a finite time can also be calculated from the occupancy of the bound surface states. From Eq. (\ref{fullrateeqn}), we infer that the losses of the bound state occupancy increase the continuum state occupancy by
\begin{align}
\frac{\mathrm{d}n_k}{\mathrm{d}t}=\sum_n W_{kn} n_n(t) \text{ .}
\end{align}
As the electron remains in the surface potential for the time it needs to travel trough the surface potential the occupancy of the continuum state \(k\) is given by \(n_k=\tau_t \dot{n_k} \). To obtain the energy-resolved desorption flux we multiply the occupancy of the continuum state \(k\) with the flux \(j_k^\text{box}\), associated with the box-normalized state \(|\phi_k\rangle\).\cite{HBF10a} Thus, the  energy-resolved desorption flux is given by
\begin{align}
j_k(t)=\tau_t j_k^\text{box} \sum_n W_{kn} n_n(t) \text{ ,} \label{eresdes}
\end{align}
which is well defined in the limit \(L\rightarrow \infty\).

\section{Transition probabilities}
\label{Electron-surface interaction and transition rates}

The kinetic equations presented in the last section rely on the knowledge of the transition probabilities. They have 
to be calculated from a microscopic model for the electron-surface interaction. 

For a dielectric surface, the transitions are driven by phonons, whose maximum energy is, within the Debye model, the 
Debye energy \(\hbar \omega_D\). Measuring energies in units of the Debye energy, important dimensionless parameters 
characterizing the potential depth are
\begin{align}
\epsilon_n=\frac{E_n}{\hbar \omega_D}  \quad \text{ and } \quad \Delta_{nn^\prime}=\frac{E_n-E_{n^\prime}}{\hbar \omega_D} \text{ ,}
\end{align}
where \(E_n<0\) is the energy of the \(\text{n}^\text{th}\) bound state. 

In I, we introduced the following classification for the potential depth. If  \(-n+1> \Delta_{12} >- n\), we call the 
potential n-phonon deep. For the calculations in I and II, we considered only one- or two-phonon deep potentials, 
for which one- and two-phonon transition probabilities are sufficient. Dielectrics with two-phonon deep potentials, 
such as graphite or MgO, are however an exception. Many dielectrics, for instance, \(\text{Al}_2\text{O}_3\), CaO, GaAs, 
or \(\text{SiO}_2\) have more than two-phonon deep potentials. Hence, the more relevant situation is physisorption in 
deep surface potentials for which multi-phonon transition probabilities are required.

To calculate multi-phonon transition probabilities for the one-dimensional microscopic model used in I and II, 
we briefly recall its main features. In short, for a dielectric surface, the main source of the attractive static 
electron-surface potential is the coupling of the electron to a dipole-active
surface phonon.\cite{EM73} Far from the surface the surface potential arising from this coupling merges with the 
classical image potential and thus \(\sim 1 / z\).
Close to the surface, however, the surface potential is strongly modified by the
recoil energy resulting from the momentum transfer parallel
to the surface when the electron absorbs or emits a surface
phonon. Taking this effect into account leads to a recoil-corrected 
image potential \(\sim 1 /(z + z_c)\) with \(z_c\) a cut-off parameter defined in I.

Transitions between the eigenstates of the recoil-corrected
image potential are due to dynamic perturbations of the surface potential. The surface potential is very steep near the
surface. A particularly strong perturbation arises therefore
from the longitudinal-acoustic bulk phonon perpendicular to the
surface which causes the surface plane to oscillate. 

The Hamiltonian from which we calculate the transition probabilities was 
introduced in I where all quantities entering the Hamiltonian are explicitly defined. It is given by
\begin{align}
H=H_{e}^\text{static}+H_{ph}+H_{e-ph}^\text{dyn} \text{ ,}
\end{align}
where 
\begin{align}
H_e^\text{static}=\sum_q E_q c_q^\dagger c_q
\end{align}
describes the electron in the recoil-corrected image potential, 
\begin{align}
H_{ph}=\sum_Q \hbar \omega_Q b_Q^\dagger b_Q 
\end{align}
describes the free dynamics of the bulk longitudinal acoustic phonon responsible for transitions between surface states, and
\begin{align}
H_{e-ph}^\text{dyn}=\sum_{q,q^\prime} \langle q^\prime | V_p(u,z) |q\rangle c_{q^\prime}^\dagger c_q
\end{align}
denotes the dynamic coupling of the electron to the bulk phonon. 

The perturbation \(V_p(u,z)\) can be identified as the difference between the displaced surface potential and the static surface potential. It reads, after the transformation \(z \rightarrow z- z_c\), 
\begin{align}
V_p(u,z)=-\frac{e^2 \Lambda_0}{z+u}+\frac{e^2\Lambda_0}{z} \label{compactperturb} \text{ ,}
\end{align}
where \(\Lambda_0=(\epsilon_s-1)/4(\epsilon_s+1)\) with \(\epsilon_s\) the static dielectric constant. 
In general, multi-phonon processes can arise both from the nonlinearity of the electron-phonon coupling 
\(H_{e-ph}^\text{dyn}\) as well as from the successive actions of \(H_{e-ph}^\text{dyn}\) encoded in the 
T matrix equation,
\begin{align}
T=H_{e-ph}^\text{dyn}+H_{e-ph}^\text{dyn} G_0 T \text{ ,}
\end{align}
where \(G_0\) is given by 
\begin{align}
G_0=\left(E-H_e^\text{static}-H_{ph}+i\epsilon\right)^{-1} \text{ .}
\end{align}
The transition probability per unit time from an electronic state \(q\) to an electronic state \(q^\prime\) 
encompassing both types of processes is given by~\cite{BY73}
\begin{align}
W_{q^\prime q}=&\frac{2\pi}{\hbar} \sum_{s,s^\prime} \frac{e^{-\beta_s E_s}}{\sum_{s^{\prime\prime}} e^{-\beta_s E_{s^{\prime \prime}}} } \left|\langle s^\prime,q^\prime | T | s,q\rangle \right|^2 \nonumber \\
& \times \delta (E_s -E_{s^\prime} +E_q -E_{q^\prime}) \text{ ,} \label{generalTR}
\end{align}
where \(\beta_s = (k_B T_s)^{-1}\),  with \(T_s\) the surface temperature and \(|s\rangle\) and \(|s^\prime \rangle\) the initial and final phonon states. We are  only interested in the transitions between electronic states. It is thus natural to average in Eq. (\ref{generalTR}) over all phonon states. The delta function guarantees energy conservation. 

In our previous work, we have used  an expansion of the T matrix to calculate multi-phonon transition rates. In principle this ensures that both linear and nonlinear terms in the interaction as well as successive actions of the perturbation are taken into account up to a certain order of the phonon process. However even for a two-phonon deep potential, taking all two-phonon processes into account is nearly impossible. The calculation becomes feasible if two-phonon processes are only taken into account for transitions not already enabled by a one-phonon process. This amounts to computing only the lowest required phonon order for a given transition, neglecting higher order corrections to it. 
For higher order phonon processes even this simplified strategy becomes unfeasible. A different approach is thus needed.

From I and II we qualitatively know the relevance of the different types of multi-phonon processes for particular
electronic transitions. For continuum-bound state transitions, for instance, one-phonon processes are sufficient at
low electron energies. We will therefore compute the transition probability between bound and continuum states
in the one-phonon approximation. For transitions between bound states, we found that multi-phonon processes due to 
the nonlinearity of the electron-phonon coupling tend to be more important than the multi-phonon processes due to 
the iteration of the T matrix, unlike to what we found for bound state - continuum transitions (see I) or to what 
\v{S}iber and Gumhalter~\cite{SG03,SG05,SG08} found in the context of atom-surface scattering. Indeed, multi-phonon
processes from the iteration of the T matrix give a minor contribution, unless resonances arising from the 
T matrix become relevant. This happens whenever the energy difference between two bound states is a multiple of 
the Debye energy. Resonances smoothen then the abrupt steps in the transition probability at the depth thresholds. 
Since the electronic matrix element between the first and the second bound state is the largest one, this effect is 
most pronounced for \(|\Delta_{12}|=n\).

In view of the above discussion we expect an approximation which takes only the nonlinearity of the electron-phonon 
interaction non-perturbatively into account to give an acceptable first estimate for the multi-phonon transition
rates. We denote this approximation the nonlinear multi-phonon approximation. In particular, it should be sufficient 
for the identification of the generic behavior of multi-phonon-mediated adsorption and desorption.

Calculating multi-phonon processes due to nonlinear terms in the interaction potential\cite{Manson91} 
amounts to a distorted wave Born approximation with the full interaction potential. Thus, the transition 
probability per unit time is given by
\begin{align}
W_{q^\prime q}=&\frac{2\pi}{\hbar} \sum_{s,s^\prime} \frac{e^{-\beta_s E_s}}{\sum_{s^{\prime \prime}} e^{-\beta_s E_{s^{\prime \prime}}}} \left|\langle q,s| H_{e-ph}^\text{dyn} |s^\prime, q^\prime \rangle\right|^2  \nonumber \\
 & \times \delta(E_s+E_q-E_{q^\prime}-E_{s^\prime})~. \label{mphTR}
\end{align}

To evaluate the multi-phonon transition probability, we use \(H_{e-ph}^\text{dyn}\) in the form of 
Eq. (\ref{compactperturb}). The transition matrix element in Eq. (\ref{mphTR}) is then given by
\begin{align}
&\langle q,s| H_{e-ph}^\text{dyn.} |q^\prime, s^\prime \rangle\nonumber \\
&\quad =\langle s| \int_{z_c}^\infty \mathrm{d}z \phi_q^\ast(z) \left[ v(z+u)-v(z)\right] \phi_{q^\prime}(z)|s^\prime \rangle \text{ ,}
\end{align} 
where \(v(z)=-(e^2\Lambda_0)/z\). Introducing dimensionless variables \(x=z/a_B\), the Fourier transform of the static potential
\begin{align}
v(p)=\int_{x_c}^{\infty} \mathrm{d}x e^{ipx}v(x)  \text{ ,}
\end{align}
and the state-to-state matrix element
\begin{align}
f_{q q^\prime}(p)=\int_{x_c}^\infty \mathrm{d}x \phi_q^\ast(x)e^{-ipx} \phi_{q^\prime}(x) ~,
\end{align}
the transition probability can be rewritten as
\begin{widetext}
\begin{align}
W_{q^\prime q}=&\frac{2\pi}{\hbar} \sum_{s,s^\prime} \frac{e^{-\beta_s E_s}}{\sum_{s^{\prime\prime}}e^{-\beta_s E_{s^{\prime\prime}}}} \int_{-\infty}^\infty \frac{\mathrm{d}p}{2\pi} \int_{-\infty}^\infty \frac{\mathrm{d}\tilde{p}}{2\pi} v(p) v^\ast(\tilde{p}) f_{q q^\prime}(p) f_{q q^\prime}^\ast(\tilde{p}) \langle s | \left[e^{-i\frac{p}{a_B}u}-1\right] |s^\prime \rangle \nonumber \\
& \times \langle s^\prime | \left[ e^{i\frac{\tilde{p}}{a_B}u} -1 \right] | s \rangle \delta(E_s+E_q-E_{s^\prime}-E_{q^\prime}) \text{ .}
\end{align}
Using the identity \(\delta (x)=1/(2\pi) \int_{-\infty}^\infty \mathrm{d}t\text{ }e^{ixt}\) and employing \(\langle s|e^{i E_s t/\hbar}=\langle s| e^{iH_{ph} t/\hbar}\), the above expression becomes
\begin{align}
W_{q^\prime q}=\frac{1}{\hbar^2}\int_{-\infty}^\infty \frac{\mathrm{d}p}{2\pi} 
\int_{-\infty}^\infty \frac{\mathrm{d}\tilde{p}}{2\pi} v(p) v^\ast(\tilde{p}) f_{q q^\prime}(p) 
f_{q q^\prime}^\ast(\tilde{p}) \int_{-\infty}^\infty \mathrm{d}t\text{ } e^{i(E_q-E_{q^\prime})t/\hbar} 
\langle\langle \left[e^{-i\frac{p}{a_B}u(0)}-1\right] \left[e^{i\frac{\tilde{p}}{a_B} u(t)} -1\right] 
\rangle \rangle
\label{Wqq}
\end{align}
with \(\langle\langle \dots \rangle \rangle = \sum_s e^{-\beta_s E_s} \langle s| \dots |s\rangle / 
\sum_{s^{\prime\prime}} e^{-\beta_s E_{s^{\prime \prime}}}\) the average over phonon states. 
This average can be evaluated for \(q\neq q^\prime\) employing Glauber's theorem\cite{Glauber55}
which yields
\begin{align}
\langle \langle \left[e^{-i\frac{p}{a_B}u(0)}-1\right] \left[e^{i\frac{\tilde{p}}{a_B}u(t)}-1\right] 
\rangle\rangle = e^{-\frac{1}{2a_B^2}p^2\langle\langle u(0)^2 \rangle \rangle } e^{-\frac{1}{2a_B^2} 
\tilde{p}^2 \langle\langle u(t)^2 \rangle \rangle} e^{\frac{1}{a_B^2}p\tilde{p} \langle\langle u(0) u(t) 
\rangle \rangle } \label{GlauberTrick}
\end{align}
with  
\begin{align}
\langle\langle u(0) u(t) \rangle \rangle = 
\sum_Q \frac{\hbar}{2 \mu N_s \omega_Q} \left[\left(1+n_B(\hbar\omega_Q)\right)e^{-i\omega_Q t} + 
n_B(\hbar\omega_Q)e^{i\omega_Q t} \right]
\end{align}
\end{widetext}
the correlation function of the displacement field
\begin{align}
u=\sum_Q \sqrt{\frac{\hbar}{2\mu\omega_Q N_s}}\left( b_Q+b_{-Q}^\dagger \right)~,
\end{align}
where \(\mu\) is the mass of the unit cell of the lattice and $N_s$ is                           
the number of unit cells.

As in I and II we use for calculational convenience a bulk Debye model for the longitudinal acoustic phonon, although it is less justified for the high energy part of the spectrum  which also enters our calculation. Sums over phonon momenta are thus replaced by
\begin{align}
\sum_Q \dots = \frac{3 N_s}{\omega_D^3}\int_0^{\omega_D} \mathrm{d}\omega \text{ } \omega^2 \dots \text{ .}
\end{align}
In terms of the dimensionless variables
\begin{align}
x=\frac{\omega}{\omega_D} \text{  ,  } \delta=\frac{\hbar \omega_D}{k_B T_s} \text{  , and } \tau=\omega_D t \text{ ,} \label{deltadef}
\end{align}
the phonon correlation function becomes
\begin{align}
\langle \langle u(0) u(\tau) \rangle \rangle = \frac{3 \hbar}{2 \mu \omega_D}  \int_0^1 \mathrm{d}x x \left[\frac{e^{-ix\tau}}{1-e^{-\delta x}} +\frac{e^{ix\tau}}{e^{\delta x}-1} \right] \text{ .}
\end{align}
Hence, for the transition probability per unit time we obtain
\begin{widetext}
\begin{align}
W_{q^\prime q}=\frac{e^4 \Lambda_0^2}{\hbar^2 \omega_D a_B^2} \int_{-\infty}^\infty \frac{\mathrm{d}p}{2 \pi} \int_{-\infty}^\infty \frac{\mathrm{d}\tilde{p}}{2 \pi} v(p) v(\tilde{p}) f_{q q^\prime}(p) f_{q q^\prime}^\ast(\tilde{p}) e^{-\frac{1}{2} \gamma p^2 q(0)} e^{-\frac{1}{2}\gamma \tilde{p}^2 q(0)} \int_{-\infty}^{\infty} \mathrm{d}\tau e^{i\Delta_{q q^\prime}\tau +\gamma p \tilde{p} q(\tau)}  \label{fullmphrates} \text{ ,}
\end{align}
where 
\begin{align}
q(\tau)= \int_0^1 \mathrm{d}x  x \left[ \frac{e^{-ix\tau}}{1-e^{-\delta x}}+\frac{e^{ix\tau}}{e^{\delta x}-1} \right] \qquad \text{and} \qquad \gamma = \frac{3 \hbar}{2\mu a_B^2 \omega_D} \text{ .} 
\end{align}
\end{widetext}

The transition probability (\ref{fullmphrates}) contains two Debye-Waller factors, 
\(\exp(-\gamma p^2 q(0)/2)\) and \(\exp(-\gamma \tilde{p}^2 q(0)/2)\), governing the reduction of the 
transition probability as a function of the surface temperature. It also contains phonon processes of 
all orders as can be most easily seen from the Taylor expansion
\begin{align}
e^{\gamma p \tilde{p} q(\tau)}=1+\gamma p \tilde{p}q(\tau) +\frac{1}{2} \left[ \gamma p \tilde{p} q(\tau) \right]^2
+\dots~. \label{mphexpansion}
\end{align}
Clearly, the second term on the right hand side represents the one-phonon and the third term the two-phonon process.
From I we know that two-phonon processes are much weaker than one-phonon processes. We expect 
therefore lower order phonon processes to dominate their higher order corrections, so that the expansion
(\ref{mphexpansion}) converges quickly. 

As higher order phonon processes are small compared to lower order processes, we take, for a given 
\(\Delta_{qq^\prime} \) only the leading term of \(\exp( \gamma p \tilde{p} q(\tau)) \) into account, 
that is, the lowest order phonon process that enables a transition between the states \(q\) and \(q^\prime\). The 
Fourier transformation of powers of \(q(\tau)\), however, required when (\ref{mphexpansion}) is used in 
(\ref{fullmphrates}), cannot be evaluated in closed from, making it necessary to construct an approximation for \(q(\tau)\). 

To derive an approximation for \(q(\tau)\) subject to the constraint 
\begin{align}
\int_{-\infty}^\infty \mathrm{d}\tau e^{i\Delta_{qq^\prime} \tau} q^n(\tau)=0 \quad \text{ for } |\Delta_{qq^\prime}|>n 
\text{ ,} \label{vanishcondition}
\end{align}
which states that an n-phonon process yields a non-vanishing transition probability only for \(-n<\Delta<n\) and 
vanishes otherwise, we split \(q(\tau)=q^s(\tau)+q^i(\tau)\) into a contribution arising from 
spontaneous phonon emission \(q^s(\tau)\) and a contribution from induced phonon emission or absorption \(q^i(\tau)\).
They are respectively given by
\begin{align}
q^s(\tau)=\int_0^1 \mathrm{d}x \text{ }xe^{-ix\tau} \text{ and } q^i(\tau)=2 \int_0^1 \mathrm{d}x\text{ }x \frac{\cos(x\tau)}{e^{\delta x}-1} \text{ .}
\end{align}
The former can be evaluated giving
\begin{align}
q^s(\tau)=\frac{\cos \tau -1}{\tau^2}+i\frac{\tau \cos \tau- \sin \tau}{\tau^2} +\frac{\sin \tau}{\tau} \text{ .}
\end{align}
For \(q^i(\tau)\) we need to find an approximation. For that purpose we look at the Fourier transform of \(q^i\)
\begin{align}
\int_{-\infty}^\infty \mathrm{d}\tau e^{i \Delta \tau} q^i(\tau) =\left\lbrace 
\begin{matrix}
2\pi \frac{|\Delta|}{e^{\delta |\Delta|}-1}  & \text{for} -1 <\Delta<1 \\
 0 & \text{else}
\end{matrix}
 \right.~.
\end{align}
Expanding the Fourier transform in terms of \(|\Delta|\), 
\begin{align}
2\pi \frac{|\Delta|}{e^{\delta |\Delta|}-1} \approx 2 \pi \left[\frac{1}{\delta}-\frac{1}{2}|\Delta| 
+\frac{1}{12}\delta |\Delta|^2+\mathcal{O}(\delta^2) \right] ~,
\end{align}
yields a high-temperature approximation which converges quickly for the temperatures we are interested in 
and guarantees at the same time that the one-phonon contribution can only bridge energy differences up to 
\(|\Delta_{qq^\prime}|=1\). Applying the inverse transformation gives
\begin{align}
q^i(\tau)=&\left(\frac{2}{\delta}-1+\frac{\delta}{6} \right)\frac{\sin(\tau)}{\tau}+\frac{1}{\tau^2} \nonumber \\
&+\left(-1+\frac{\delta}{3}\right) \frac{\cos(\tau)}{\tau^2}-\frac{\delta}{3}\frac{\sin(\tau)}{\tau^3}+\mathcal{O}(\delta^2) \text{ ,} \label{htexpans}
\end{align}
which satisfies Eq. (\ref{vanishcondition}). 
Using this approximation the Fourier transform of powers of \(q(\tau)\) can be done analytically. 

As the \(n\)-phonon process gives a vanishing transition probability at \(|\Delta|=n\), we take the maximum of the \(n\)-phonon and the \(n+1\)-phonon process to obtain a better approximation in the vicinity of  \(|\Delta|=n\). Then the Fourier transformation of \(\exp(\gamma p \tilde{p} q(\tau))\) in leading non-vanishing order is given by 

\begin{widetext}

\begin{align}
\int_{-\infty}^\infty \mathrm{d}\tau e^{i \Delta \tau +\gamma p \tilde{p} q(\tau,\delta)}\approx\left\lbrace
\begin{matrix}
\max (A_n,A_{n+1}) &  \text{ for } n-1 < \Delta <n \\
\max (B_n,B_{n+1}) & \text{ for } -n < \Delta <-n+1 
\end{matrix} \right. \text{ ,} \label{expqap}
\end{align}
where
\begin{align}
A_n=- 2\pi \frac{\left(\gamma p \tilde{p}\right)^n}{n!} \sum_{k=0}^n {n \choose k} \sum_{j=0}^k { k \choose j} \left(-1\right)^{n+j} \left(\frac{1}{\delta}+\frac{1}{2}+\frac{\delta}{12}\right)^{n-k}\left(\frac{1}{2}+\frac{\delta}{6} \right)^{k-j}\left(-\frac{\delta}{6} \right)^j \frac{\left(\Delta-n\right)^{n+k+j-1}}{\left(n+k+j-1\right)!}+\mathcal{O}(\delta^2)
\end{align}
and
\begin{align}
B_n=2\pi \frac{\left(\gamma p \tilde{p}\right)^n}{n!} \sum_{k=0}^n {n \choose k} \sum_{j=0}^k {k \choose j} \left(-1\right)^{n+j} \left(-\frac{1}{\delta}+\frac{1}{2}-\frac{\delta}{12}\right)^{n-k}\left(-\frac{1}{2}+\frac{\delta}{6}\right)^{k-j} \left(\frac{\delta}{6}\right)^j  \frac{\left(\Delta+n \right)^{n+k+j-1}}{\left(n+k+j-1\right)!}+ \mathcal{O}(\delta^2) \text{ .} \label{Bn}
\end{align}

\end{widetext}

Using the approximation given by Eq. (\ref{expqap}) allows an efficient numerical evaluation of the transition probabilities  (\ref{fullmphrates}).

Equations (\ref{htexpans}-\ref{Bn}) are first order in \(\delta\). For the materials and temperatures we are interested in
this is sufficient. Note, however, that the expansion can be continued to higher orders in \(\delta\). The 
Fourier transformation of \(\exp(\gamma p \tilde{p} q(\tau) )\) is then still a polynomial in \(\Delta\) and thus 
amenable for numerical calculations.

\section{Results}
\label{Results}

We now use the multi-phonon transition probability to study the electron kinetics in front of a 
\(\text{CaO}\), \(\text{Al}_2\text{O}_3\), and \(\text{SiO}_2\) surface. They all have three-phonon deep surface
potentials, that is, the energy difference of the two lowest image states is between two and three
Debye energies. The material parameters required for the numerical computation are summarized in Table \ref{materialtable}. 
All numerical results were obtained for these parameters. Where indicated, we varied the Debye temperature to simulate 
different potential depths. Furthermore, the multi-phonon calculation for \(\text{CaO}\), \(\text{Al}_2\text{O}_3\), and 
\(\text{SiO}_2\) is compared to the one- and two-phonon calculations from I and II which are applicable to graphite and 
MgO.
\begin{table}[t]
\caption{Material parameters for the numerical results.   }
\center
\begin{tabular}{l l l l l}
 &  & CaO & \(\text{Al}_2\text{O}_3\) & \(\text{SiO}_2\) \\
\hline
Debye temperature  \(T_D\) & \(\quad\) & \(648\) K &\(980\) K &\(470\) K  \\
Dielectric constant  \(\epsilon_s\) & \(\quad\) &\(12.01\) &\(9.9\) & \(3.78\) \\
TO-phonon  frequency  \(\hbar\omega_T\) & \(\quad\) & \(41\) meV  & \(79\) meV  & \(133\) meV \\
\hline
\end{tabular}
\label{materialtable}
\end{table}

\subsection{Desorption}

\begin{figure}
\includegraphics[width=\linewidth]{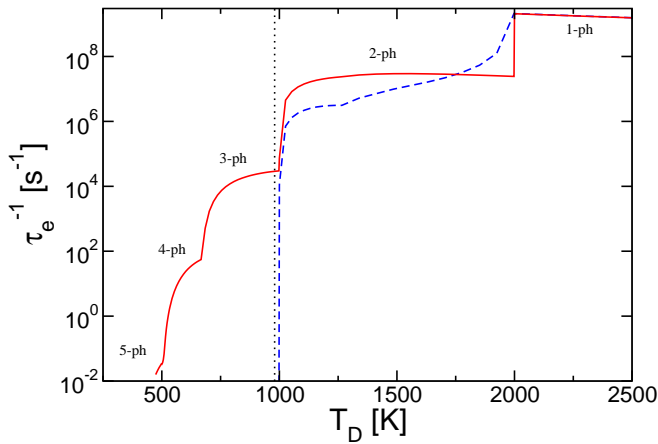}
\caption{(Color online) Inverse desorption time \(\tau_e^{-1}\) as a function of the Debye temperature \(T_D\) 
for \(\delta=1\) (surface temperature \(T_s=T_D/\delta\)) calculated in the nonlinear multi-phonon approximation 
(solid red line) and the
two-phonon approximation from I for two-phonon depth (dashed blue line). The surface potential is one-phonon deep for \(T_D >2000K \), two-phonon deep for \( 2000K> T_D >1000K \), three-phonon deep for \( 1000K> T_D >666K \) and four-phonon deep for \(666K > T_D >500K \). Data for \(T_D=980K\) apply to \(\text{Al}_2\text{O}_3\) (thin vertical line).}
\label{figure1}
\end{figure}

To judge the quality of the nonlinear multi-phonon approximation derived in the previous section, 
we first compare in Fig.~\ref{figure1} the inverse desorption time obtained from it with the 
inverse desorption time obtained from our previous one- and two-phonon approximation.
Shown is the dependence of \(\tau_e^{-1}\) on the Debye temperature \(T_D\) which is 
tuned to vary the potential depth. The dimensionless inverse temperature \(\delta\), as defined 
in Eq. (\ref{deltadef}), is kept constant to keep the level of phonon excitation the same while 
the Debye temperature is varied. 

The nonlinear multi-phonon approximation can be of course only compared with the two-phonon 
approximation in the range of Debye temperatures for which the potential is two-phonon deep. 
Calculated in the multi-phonon approximation \(\tau_e^{-1}\) changes very little over the range of 
two-phonon depth, but shows steep jumps at the threshold to one- and three-phonon depth. For \(\tau_e^{-1}\) 
calculated in the two-phonon approximation, which is based on an iteration of the T matrix with the nonlinear 
electron-phonon coupling, these thresholds are washed out by the resonances. Nevertheless \(\tau_e^{-1}\) is 
on the same order of magnitude in both approximations. The main effect of the neglected resonances is the 
rounding off of the drops at the thresholds. As the steps, an artefact of taking only nonlinear multi-phonon 
processes into account, are less steep for deeper potentials, the nonlinear multi-phonon approximation might 
be even more appropriate for deeper potentials. The thin dotted vertical line in Fig.~\ref{figure1} corresponds 
to \(\text{Al}_2\text{O}_3\). Unfortunately the potential depth is just below the two-phonon three-phonon threshold, 
so that the value for \(\tau_e^{-1}\) is most likely underestimated.

\begin{figure}
\includegraphics[width=\linewidth]{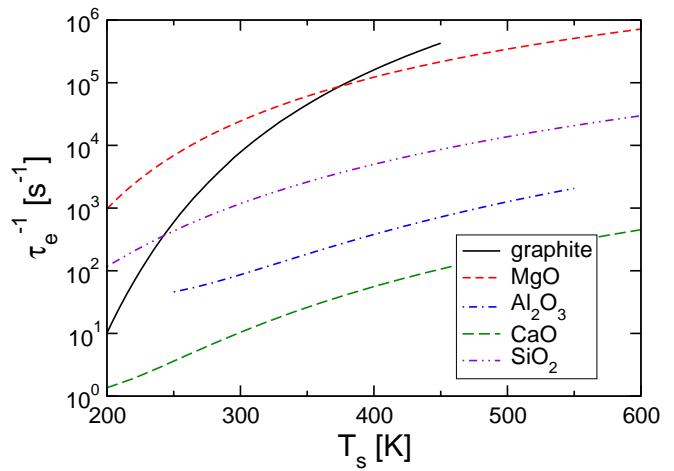}
\caption{(Color online) Inverse desorption time \(\tau_e^{-1}\) as a function of the surface temperature \(T_s\) for graphite, MgO, \(\text{CaO}\), \(\text{Al}_2\text{O}_3\), and \(\text{SiO}_2\). }
\label{figure2}
\end{figure}

We now move on to the study of the dependence of \(\tau_e^{-1}\) on the surface temperature. Figure~\ref{figure2} 
shows the inverse desorption time \(\tau_e^{-1}\) as a function of the surface temperature for graphite, MgO, 
\(\text{Al}_2\text{O}_3\), CaO, and \(\text{SiO}_2\). For graphite and MgO, both two-phonon deep, \(\tau_e^{-1}\) 
was calculated in the two-phonon approximation, for \(\text{Al}_2\text{O}_3\), CaO and \(\text{SiO}_2\), all of 
them three-phonon deep, the nonlinear multi-phonon approximation has been used. For all materials \(\tau_e^{-1}\) 
increases significantly with the surface temperature. 

Comparing in Fig.~\ref{figure2} \(\tau_e^{-1}\) for \(\text{Al}_2\text{O}_3\), CaO, and \(\text{SiO}_2\), we notice
that \(\tau_e^{-1}\) increases with decreasing \(\epsilon_s\) (see Table \ref{materialtable}) in accordance with the 
fact that a smaller \(\epsilon_s\) implies a less deep surface potential and thus a faster desorption. From 
Fig.~\ref{figure2} we also see that for high surface temperatures desorption from the two-phonon deep potentials of graphite 
and MgO is quicker than from the three-phonon deep potentials of \(\text{Al}_2\text{O}_3\), CaO and \(\text{SiO}_2\) as 
expected. For low surface temperature, however, \(\tau_{e}^{-1}\) for graphite decreases much steeper than for the other 
materials. This might be due to the high Debye temperature of graphite, so that for room temperature the dimensionless 
inverse temperature \(\delta=T_D/T_s\) which controls phonon excitation is already in the low temperature regime where 
downwards transitions due to spontaneous phonon emission remain constant while upwards transitions are extremely 
temperature dependent, causing the desorption time to be equally temperature dependent. This peculiarity leads
to the surprising fact that at low temperatures desorption from two-phonon deep potentials can be in some cases
slower than desorption from three-phonon deep potentials.

Figure~\ref{figure2} also gives insight into the validity of the high temperature expansion (\ref{htexpans}). Up to 
first order in \(\delta\), it is valid for \(\delta < 3\). This corresponds to a surface temperature of 300K for 
\(\text{Al}_2\text{O}_3\) and 225K for CaO. The small upwards bends at these temperatures indicate that for lower 
surface temperatures the expansion given by Eq. (\ref{htexpans}) should be continued to higher orders in \(\delta\).

\subsection{Sticking}

In II we found that one-phonon processes give much higher contributions to the sticking coefficient than two-phonon 
processes. For this reason we calculate the transition probabilities for continuum-bound state transitions only in 
the one-phonon approximation. The effect of multi-phonon processes with regard to sticking lies in the relaxation 
from the state the electron is initially trapped to the lowest bound state. This is captured by the kinetic sticking 
coefficient. Before we address this question in more detail we take a look at the prompt sticking
coefficient. 

\begin{figure}
\includegraphics[width=\linewidth]{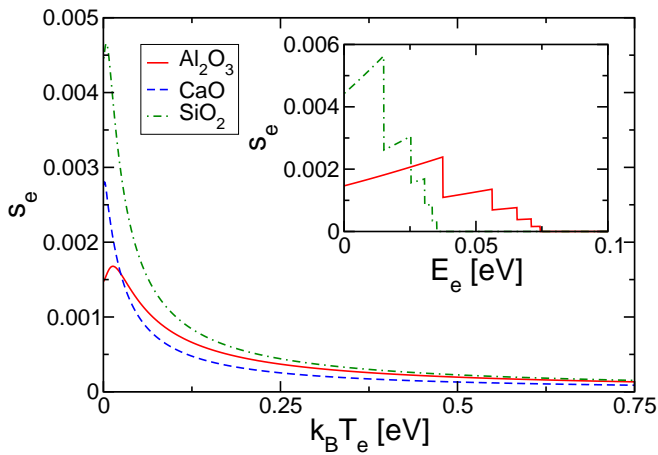}
\caption{(Color online) Prompt energy-averaged sticking coefficient for  \(\text{CaO}\), \(\text{Al}_2\text{O}_3\), and \(\text{SiO}_2\) as a function of the mean energy of the electron at a surface temperature of \(T_s=300 K\). Inset: Prompt energy-resolved sticking coefficient for  \(\text{Al}_2\text{O}_3\) and \(\text{SiO}_2\) as a function of the electron energy for \(T_s=300K\).}
\label{figure3}
\end{figure}

The prompt sticking coefficient for \(\text{Al}_2\text{O}_3\), CaO and \(\text{SiO}_2\) is presented in 
Fig.~\ref{figure3}. First we consider the prompt energy-resolved sticking coefficient shown in the inset. Note that 
the quadratic phonon dispersion of the Debye model translates into an energy-resolved sticking coefficient which, 
apart from the discontinuities, is proportional to the electron energy. The steep jumps in the energy-resolved sticking 
coefficient reflect level accessibility. When the energy difference between the approaching electron and a bound state 
exceeds the Debye energy, one-phonon processes no longer enable sticking to this level. Since the lowest two bound 
states of the image potential of \(\text{Al}_2\text{O}_3\) and \(\text{SiO}_2\) have energies \(\epsilon_n <-1\), they 
cannot be reached by one-phonon processes from the continuum. Thus, the lowest bound state contributing to prompt 
sticking is the third bound state. Due to the differences in the Debye energy, the energy-resolved sticking coefficient 
for \(\text{SiO}_2\) is larger and increases faster for low electron energies than the sticking coefficients for 
\(\text{Al}_2\text{O}_3\) and CaO (not explicitly shown). Compared to \(\text{Al}_2\text{O}_3\) and  CaO the 
energy-resolved sticking coefficient for \(\text{SiO}_2\) is thus strongly peaked at low electron energies. As a result, 
the energy averaged-sticking coefficient shown in the main panel of Fig.~\ref{figure3} is much larger for \(\text{SiO}_2\) 
than for \(\text{Al}_2\text{O}_3\) and CaO. 

\begin{figure}
\includegraphics[width=\linewidth]{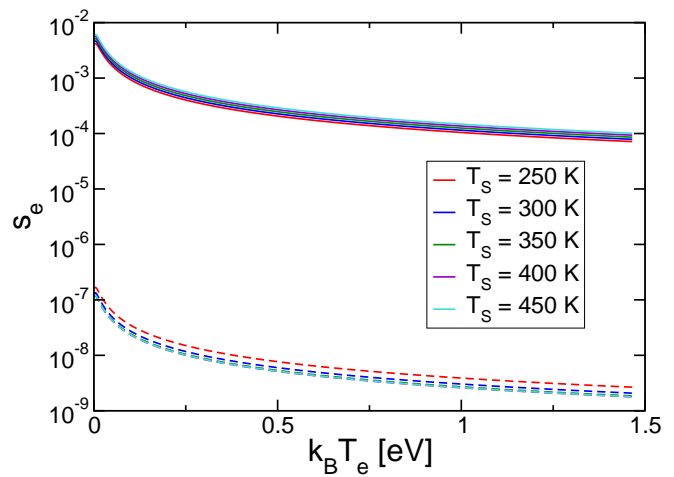}
\caption{(Color online) Prompt (full line) and kinetic (dashed line) energy averaged sticking coefficient for \(\text{SiO}_2\) as a function of the mean energy of the electron and the surface temperature \(T_s\).}
\label{figure4}
\end{figure}

Figure~\ref{figure4} shows the prompt and kinetic energy averaged sticking coefficient for \(\text{SiO}_2\) as a 
function of the mean electron energy and the surface temperature. The prompt sticking coefficient increases slightly 
with temperature due to the increased contribution of induced phonon emission responsible for continuum-bound state 
transitions. The kinetic sticking coefficient is smaller than the prompt sticking coefficient by four to five orders
of magnitude and decreases with temperature as a higher surface temperature favors quick transitions back into the 
continuum after initial trapping. 

Depending on whether transitions from the upper bound states to the lowest state or to the continuum are more likely, 
the electron trickles through after initial trapping or desorbs before relaxing to the lowest bound state. For the 
three-phonon deep surface potentials of \(\text{Al}_2\text{O}_3\), CaO and \(\text{SiO}_2\) trickling through is 
suppressed leading to a considerable reduction of the kinetic compared to the prompt sticking coefficient. 

\subsection{Electron kinetics}

So far we have calculated from the kinetic rate equation (\ref{fullrateeqn2}) the prompt and kinetic sticking coefficients and the desorption time. The rate equation contains however more information. For a specified electron influx or initial condition, the time-evolution of the bound state occupancy and the energy resolution of the desorption flux can be calculated as well. 

\begin{figure}
\includegraphics[width=\linewidth]{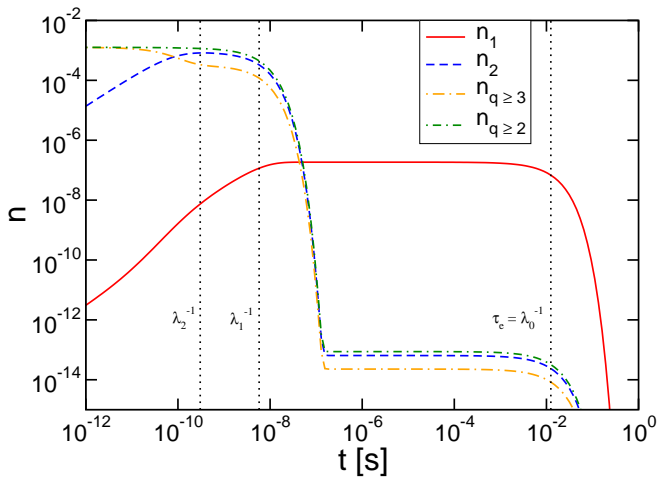}
\caption{(Color online) Time evolution of the bound state occupancy of a single electron trapped at \(t=0\) in the upper bound states of \(\text{Al}_2\text{O}_3\) at \(T_s=300K\). The thin vertical lines correspond to \(\lambda_0^{-1}\), \(\lambda_1^{-1}\), and \(\lambda_2^{-1}\), respectively, where \(-\lambda_i\) are the three lowest eigenvalues of the matrix \(\mathbf{T}\).}
\label{figure5}
\end{figure}

To address the first question, we plot in Fig.~\ref{figure5} the time evolution of the bound state occupancy. Our aim is to identify the stages of physisorption and to relate them to the eigenvalues of the matrix \(T_{nm}\). The situation we are considering is a three-phonon deep potential, that is, \(-3<\Delta_{12} < -2\), with the second bound state lying more than one Debye energy below the continuum, that is, \(\epsilon_2 <-1\), so that electron trapping, due to one-phonon processes, can occur only in the third and higher bound states. The bound states \(i\ge 2\), which we call upper bound states, are linked by one-phonon transitions. 

To obtain the time evolution of the bound state occupancy after trapping of an electron at \(t=0\), we solve the rate equation with the initial condition for the bound state occupancy, \(n_i(0)=\tau_t W_{ik}\), for a specified \(k\), which is the probability that the electron is trapped in state \(i\). In Fig.~\ref{figure5}, \(k\) corresponds to an electron energy of \(E=0.05 eV\). Due to the high electron energy the third state cannot be reached by a one-phonon process. Thus trapping occurs in the fourth and higher bound states.

In a first stage after trapping of the electron, the fast one-phonon transitions between the upper bound states dominate the electron kinetics. Due to trapping of the electron in the fourth and higher bound states, the occupancy of the upper bound states is out of equilibrium.  
Over the time scale set by \(\lambda_2^{-1}\), the inverse of the third eigenvalue of \(\mathbf{T}\), the occupancy in the upper bound states relaxes towards its equilibrium value. 
The electron trickles through from the fourth and higher bound states to the second bound state, as can be seen from the increase in the occupancy of the second bound state \(n_2\) and the reduction of the occupancy of the third and higher bound states \(n_{q\ge3}\). 

Then the strong one-phonon transitions between the upper bound states and the continuum, occurring over the time scale set by \(\lambda_1^{-1}\), empty the upper bound states. The weak multi-phonon transitions from the upper states to the lowest bound state are only a small perturbation to the electron kinetics in the upper bound states so that  \(\lambda_1^{-1}\) corresponds to the desorption time for the system of the upper bound states without the lowest bound state. 

Until the upper bound states are emptied, a small fraction of the occupancy reaches the lowest bound state as can be seen from the discrepancy of the initial occupancy of the upper bound states \(n_{q \ge 2}\) and the maximum occupancy of \(n_1\). This difference corresponds to the reduction of the kinetic with respect to the prompt sticking coefficient in Fig.~\ref{figure4}. 
The lowest bound state remains occupied for a much longer time, until desorption takes place at times on the order of \(\tau_{e}=\lambda_0^{-1}\).

\begin{figure}
\includegraphics[width=\linewidth]{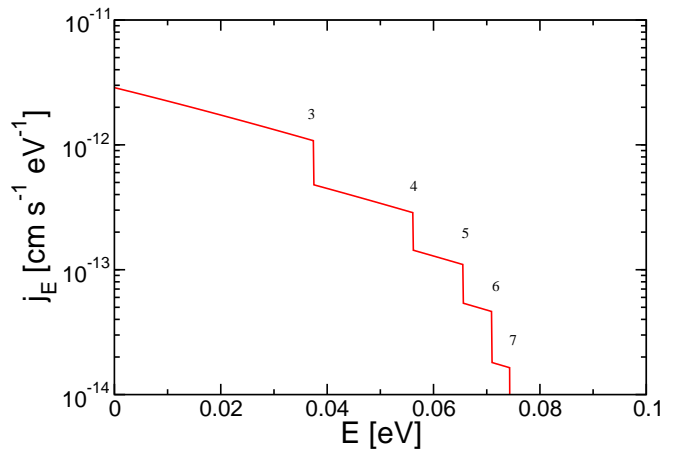}
\caption{(Color online) Energy-resolved desorption flux at \(t=10^{-4}s\) for an electron tapped at \(t=0\) in the upper bound states 
under the same conditions as in Fig.~\ref{figure5}. For \(\text{Al}_2\text{O}_3\) one-phonon transitions between bound 
and continuum states are only possible from the third and higher bound state. The small numbers give the bound state 
from which transitions to the continuum are no longer possible at the respective energy.}
\label{figure6}
\end{figure}

Figure~\ref{figure6} finally shows the energy-resolved desorption flux at \(t=10^{-4} s\)  (given by Eq. (\ref{eresdes})) after trapping of the electron under the same conditions as in Fig.~\ref{figure5}. The final transition that sets the electron free is a one-phonon transition from one of the upper bound states to the continuum. From each bound  state \(i\), one-phonon transitions are only possible to continuum states with an energy  \(E\le E_i+\hbar \omega_D\). Hence, the energy-resolved desorption flux exhibits the same discontinuities as the energy-resolved sticking coefficient shown Fig.~\ref{figure3}, located at electron energies for which one-phonon transitions between bound states and the continuum cease to be operational. 

\section{Two-state system: Discussion}
\label{Discussion: Two  state system}

To clarify the generic behavior of electron physisorption at dielectric surfaces, and to put the
results presented in the previous section and in I and II into perspective, we study a simple model of
two bound states coupled to a continuum of states. Electron physisorption occurs in the image potential 
which supports a deep lowest bound state, well separated from a group of relatively closely packed 
upper bound states. Since the upper bound states are strongly coupled by one-phonon processes they can 
be subsumed under an effective upper bound state. The effective state is then weakly coupled to the 
lowest bound state via multi-phonon processes and strongly coupled to the continuum via one-phonon
processes.

\begin{figure}
\includegraphics[width=\linewidth]{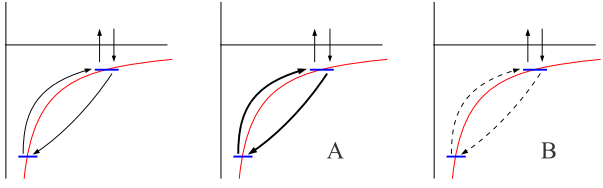}
\caption{(Color online) Left panel: Schematic drawing of the two-state model discussed in the main text. Middle panel: 
Physisorption scenario of type A. A trapped electron has a high chance to drop to the bottom, then it 
revolves between the two bound states, until it desorbs. Right panel: Physisorption scenario of type B. 
Due to a relaxation bottleneck the electron is unlikely to drop to the lowest state. Transitions that 
the electron makes once per temporary trapping event are represented by a thin line, a bold line 
represents transitions made more than once, and dashed lines represent transitions that are made with 
a very low probability.}
\label{figure7}
\end{figure}

The left panel of Fig.~\ref{figure7} schematically shows the system of the two surface states. Gearing towards
deep potentials, we include only transitions between the two bound states and between the upper state and the
continuum. The matrix \(\mathbf{T}\) defined by Eq. (\ref{fullrateeqn}) reads for this system
\begin{align}
\mathbf{T}=
\begin{pmatrix}
-W_{21} & W_{12} \\
W_{21} & -W_{12}-W_{c2}
\end{pmatrix} \text{ ,}
\end{align}
where \(W_{12}\) and \(W_{21}\) are the transition probabilities from the second to the first bound state and
vice versa, and \(W_{c2}\) is the transition probability from the second bound state to the continuum. For the
two-state model the eigenvalues \(-\lambda_{\kappa}\) and the right and left eigenvectors \(\mathbf{e}^{(\kappa)}\)
and \(\tilde{\mathbf{e}}^{(\kappa)}\) can be calculated analytically. The eigenvalues are given by
\begin{align}
-\lambda_{0,1}=&-\frac{1}{2}\left(W_{21}+W_{12}+W_{c2}\right) \nonumber \\
 & \pm \frac{1}{2}\sqrt{\left(W_{21}+W_{12}+W_{c2}\right)^2-4 W_{21} W_{c2}} \text{ .}
\end{align}

Parameters of physical interest can be also obtained analytically. The desorption time, for 
instance, is the negative inverse of the lowest eigenvalue, \(\tau_e=\lambda_0^{-1}\). 
Since only the upper bound state can be 
reached from the continuum, prompt sticking arises solely from trapping in the upper bound state. The prompt 
sticking coefficient is thus given by \(s_{e,k}^\text{prompt}=\tau_t W_{2 k}\). In the two-state model, the 
kinetic sticking coefficient is moreover related to the prompt sticking
coefficient by \(s_e^\text{kin}=\tilde{e}_2^{(0)}s_e^\text{prompt}\). Hence, the probability for the
electron to trickle through from the upper to the lower bound state is \(\tilde{e}_2^{(0)}\).

For many dielectrics the weakest transitions are from the lowest bound state to the upper bound states. 
They are typically triggered by more than two phonons. To mimic this situation within the two-state model
we set \(W_{21} \ll W_{12},W_{c2}\). The inverse of the desorption 
time becomes in this limit 
\begin{align}
\tau_e^{-1}=\frac{W_{c2}}{W_{12}+W_{c2}}W_{21} \label{2statel}
\end{align}
and the ratio between kinetic and prompt sticking coefficient becomes
\begin{align}
\frac{s_e^\text{kin}}{s_e^\text{prompt}}=\frac{W_{12}}{W_{12}+W_{c2}} \text{ .} \label{2statee}
\end{align}

The physical behavior of the two-state model depends therefore on the ratio between 
\(W_{12}\) and \(W_{c2}\) and thus on the potential depth and the surface temperature.  Two extreme 
cases are possible and represent different physisorption scenarios. For \(W_{c2} \ll W_{12}\), 
\(\tau_e^{-1}\cong (W_{21}/W_{12}) W_{c2}\), which, using detailed balance, can be brought 
into the Arrhenius form \(\tau_e^{-1}=e^{-\beta(E_2-E_1)}W_{c2}\). Kinetic and prompt sticking coefficients
coincide moreover in this parameter range. Hence, an electron trapped in the upper state drops to the 
lowest state before desorption. Desorption from the lowest state occurs then via a cascade, that is, a 
series of fast transitions \(1 \rightarrow 2 \rightarrow 1 \rightarrow 2 \rightarrow 1 \dots\) until eventually 
the transition \(2\rightarrow \text{cont.}\) removes the electron from the bound states. The just described 
physisorption scenario, which we call type A scenario, is illustrated in the middle panel of \ref{figure7}.
Recalling that the upper level stands for a manifold of strongly coupled bound states, it resembles 
the physisorption of neutral particles via cascades originally proposed and investigated by Gortel and 
coworkers.~\cite{GKT80a}

In the other limit, \(W_{12} \ll W_{c2}\). The inverse of the desorption time and the ratio between prompt 
and kinetic sticking coefficient are then given by \(\tau_e^{-1}\cong W_{21}\) and 
\(s_e^\text{kin}/s_e^\text{prompt}\cong W_{12}/W_{c2}\), respectively. The physisorption scenario is now 
dramatically different
from the one discussed before because the desorption time is solely determined by the transition probability 
from the lower to the upper bound state. As a result, desorption does not occur via a cascade, but as  
a one-way process \(1\rightarrow 2 \rightarrow \text{cont.}\), where the second transition is so fast that 
it basically does not affect the desorption time. Hence, in this scenario, which we call type B, the upper 
bound state can be considered as de-facto belonging to the continuum and desorption as basically 
equivalent to desorption from a single deep state. For sticking, the type B scenario exhibits moreover a 
relaxation bottleneck. An electron trapped in the upper state is very unlikely to drop to the lowest bound 
state, as schematically shown in the right panel of Fig.~\ref{figure7}.

Within the limits set by the model for the electron-surface interaction introduced in I and briefly recalled 
in section \ref{Electron-surface interaction and transition rates}, the two-state model contains the essential 
physics of electron physisorption. For potentials with \(\epsilon_2>-1\), where a direct one-phonon transition 
from the second bound state to the continuum is possible the two-state model can be applied directly.
Calculating the desorption time within the two-state model shows very good agreement with the results for 
graphite obtained in I. For the ratio between kinetic and prompt sticking coefficient 
\(s^\text{kin} / s^\text{prompt}\), which is 
given in the two-state model by \(\tilde{e}_2^{(0)}\), the agreement is less good but qualitatively correct, 
reproducing, for instance, the temperature dependent transition between type A and type B.
For potentials with \(\epsilon_2<-1\) no one-phonon process from the second bound state to the continuum 
is possible and the two-state model cannot be applied directly. For physisorption of type B, however, the 
electron kinetics in the upper bound states is only marginally perturbed by transitions to and from the 
lowest bound state. The time it takes an electron to get from the second bound state into the continuum is 
then the desorption time of the system of the upper bound states alone, that is, the negative inverse of the 
smallest eigenvalue \(-\lambda_0^\text{up}\) of the matrix \(T_{nm}^\text{up}\) which is the matrix 
\(T_{nm}\) defined in Eq. (\ref{fullrateeqn2}) with \(n,m > 1\). In the two-state model \(\lambda_0^\text{up}\) 
can be regarded as an effective transition rate between the second state and the continuum. Hence, to apply 
the two-state model with potentials where the second bound state does not couple by one-phonon processes to 
the continuum we simply replace in Eqs. (\ref{2statel}) and (\ref{2statee}) \(W_{c2}\) by \(\lambda_0^\text{up}\).

Let us finally look at the results obtained in the previous section and in I and II from the 
perspective of the two-state model. In a one-phonon deep potential, the transitions from the upper bound 
states to the lowest bound state and from the upper bound states to the continuum are enabled by one-phonon 
processes. In this case the downward transitions are always more likely than the upward transitions so that 
one-phonon deep potentials give always rise to physisorption of type A. Hence, they show no relaxation bottleneck
and prompt and kinetic sticking coefficient coincide. Two- or more-phonon deep potentials can either lead to 
physisorption of type A or type B, depending on the surface temperature. In 
this case one-phonon transitions from the upper bound states to the continuum compete with multi-phonon transitions 
from the upper bound states to the lowest bound state. As a transition from the upper states to the continuum requires
phonon absorption, proportional to \(n_B\), while a transition from the upper state to the lower requires phonon emission, 
proportional to \(1+n_B\), we expect that for sufficiently low temperature physisorption is always of type A, even for 
multi-phonon deep potentials.
For sufficiently high temperatures all two- or more-phonon deep potentials are however of type B. In this case a 
relaxation bottleneck results in the discrepancy between prompt and kinetic sticking coefficient (see Fig.~\ref{figure4}). 
The electron kinetics is primarily determined by the one-phonon transitions among the upper states (see Fig~\ref{figure5}). 
The temperature at which type A merges into type B depends on the potential depth and the Debye temperature.
For room temperature the three-phonon deep potentials of \(\text{Al}_2\text{O}_3\), CaO and \(\text{SiO}_2\) and the 
two-phonon deep potential of MgO are all of type B. The crossover between type A and type B occurs for the two-phonon 
deep potential of graphite at room temperature (see Fig.~5 of our previous work II). 
%From our findings we expect that for room temperature 
%all three or more phonon deep potentials are of type B and only two-phonon deep potentials can be of type A or at the 
%merger as, for instance, graphite. 

\section{Conclusions}
\label{Summary}

Within a simplified one-dimensional model for the polarization-induced interaction between an external electron 
and a dielectric surface with a sufficiently large energy gap and a sufficiently negative electron affinity, 
we investigated phonon-induced adsorption and desorption of an electron at a dielectric surface. The required
electron energy relaxation, inducing transitions between the eigenstates of the surface potential, 
which we approximated by a recoil-corrected image potential, is due to the coupling to an acoustic 
bulk phonon. 

The majority of dielectrics of interest have a surface potential that is three- or more-phonon deep, that is, 
the energy difference between the two lowest bound states is more than two Debye energies of the bulk phonon. 
In our previous work,~\cite{HBF10a,HBF10b} we took multi-phonon processes into account 
using a T matrix approach, which is however only feasible for one- and two-phonon deep potentials, as it 
is, for instance, the case for graphite. To overcome this limitation
we derived in this work a non-perturbative expression for the multi-phonon transition probability  
arising solely from the nonlinearity of the electron-phonon interaction. In view of our previous 
results for one- and two-phonon deep potentials we expect this approximation to give an acceptable 
correct order of magnitude estimate for the multi-phonon transition probability involving more than 
two phonons, despite the neglect of resonant processes stemming from the iteration of the T matrix. 

We presented numerical results for the electron desorption time for graphite, MgO, CaO, \(\text{Al}_2\text{O}_3\),
and \(\text{SiO}_2\) and the prompt and kinetic energy-resolved and energy-averaged electron sticking coefficient 
for CaO, \(\text{Al}_2\text{O}_3\), and \(\text{SiO}_2\). In addition, we calculated the energy-resolved desorption 
flux and investigated the time evolution of the bound state occupancy after initial trapping of an electron, 
revealing the characteristic stages of electron physisorption: initial trapping, relaxation in the upper bound states, 
trickling through to the lowest bound state, and desorption. Ultrafast electron spectroscopy at surfaces with 
stable image states~\cite{Fauster94,HSR97,Hoefer99} should be able to resolve these different stages 
experimentally.
% and to provide additional information about the electron kinetics we discussed in this and in our 
%previous work. It would be, for instance, interesting to know the angle-distribution of the electron desorption flux,
%a quantity which is important for the complete kinetic modeling of the dilute electron gas in 
%bounded low-temperature gas discharges.

Using a simple two-state model, we finally identified two vastly different scenarios of electron physisorption,
depending on potential depth and surface temperature, and put our results, including the ones of our previous
work~\cite{HBF10a,HBF10b}, into perspective. For almost all dielectrics of practical interest, the trapped 
electron has only for very low temperatures, well below room temperature, a significant chance to trickle 
through to the lowest bound state. The desorption process in this case would then proceed via a cascade 
between the first and second bound state, until it eventually makes a transition from there to the continuum. 
The shallow bound states, albeit important for adsorption, play a minor role for desorption. The second 
bound state is the most important one. It is a relay state.  
At room temperature, however, a relaxation bottleneck prevents the trapped electron from falling to the lowest 
bound state. The electron physisorption kinetics is thus dominated by fast one-phonon transitions in the upper 
bound states. Only a small fraction of the electron trickles through to the lowest bound state and resides there 
for a very long time until it makes a one-step desorbing transition to the continuum. 

{\it Acknowledgments.}
This work was supported by the Deutsche Forschungsgemeinschaft (DFG) through the transregional 
collaborative research center SFB/TRR 24. 

%\bibliography{./ref} 

\begin{thebibliography}{37}
\expandafter\ifx\csname natexlab\endcsname\relax\def\natexlab#1{#1}\fi
\expandafter\ifx\csname bibnamefont\endcsname\relax
  \def\bibnamefont#1{#1}\fi
\expandafter\ifx\csname bibfnamefont\endcsname\relax
  \def\bibfnamefont#1{#1}\fi
\expandafter\ifx\csname citenamefont\endcsname\relax
  \def\citenamefont#1{#1}\fi
\expandafter\ifx\csname url\endcsname\relax
  \def\url#1{\texttt{#1}}\fi
\expandafter\ifx\csname urlprefix\endcsname\relax\def\urlprefix{URL }\fi
\providecommand{\bibinfo}[2]{#2}
\providecommand{\eprint}[2][]{\url{#2}}

\bibitem[{\citenamefont{Cole and Cohen}(1969)}]{CC69}
\bibinfo{author}{\bibfnamefont{M.~W.} \bibnamefont{Cole}} \bibnamefont{and}
  \bibinfo{author}{\bibfnamefont{M.~H.} \bibnamefont{Cohen}},
  \bibinfo{journal}{Phys. Rev. Lett.} \textbf{\bibinfo{volume}{23}},
  \bibinfo{pages}{1238} (\bibinfo{year}{1969}).

\bibitem[{\citenamefont{Dose et~al.}(1984)\citenamefont{Dose, Altmann,
  Goldmann, Kolac, and Rogozik}}]{DAG84}
\bibinfo{author}{\bibfnamefont{V.}~\bibnamefont{Dose}},
  \bibinfo{author}{\bibfnamefont{W.}~\bibnamefont{Altmann}},
  \bibinfo{author}{\bibfnamefont{A.}~\bibnamefont{Goldmann}},
  \bibinfo{author}{\bibfnamefont{U.}~\bibnamefont{Kolac}}, \bibnamefont{and}
  \bibinfo{author}{\bibfnamefont{J.}~\bibnamefont{Rogozik}},
  \bibinfo{journal}{Phys. Rev. Lett.} \textbf{\bibinfo{volume}{52}},
  \bibinfo{pages}{1919} (\bibinfo{year}{1984}).

\bibitem[{\citenamefont{Straub and Himpsel}(1984)}]{SH84}
\bibinfo{author}{\bibfnamefont{D.}~\bibnamefont{Straub}} \bibnamefont{and}
  \bibinfo{author}{\bibfnamefont{F.~J.} \bibnamefont{Himpsel}},
  \bibinfo{journal}{Phys. Rev. Lett.} \textbf{\bibinfo{volume}{52}},
  \bibinfo{pages}{1922} (\bibinfo{year}{1984}).

\bibitem[{\citenamefont{Fauster}(1994)}]{Fauster94}
\bibinfo{author}{\bibfnamefont{T.}~\bibnamefont{Fauster}},
  \bibinfo{journal}{Appl. Phys. A} \textbf{\bibinfo{volume}{59}},
  \bibinfo{pages}{479} (\bibinfo{year}{1994}).

\bibitem[{\citenamefont{Hoefer et~al.}(1997)\citenamefont{Hoefer, Shumay,
  Reuss, Thomann, Wallauer, and Fauster}}]{HSR97}
\bibinfo{author}{\bibfnamefont{U.}~\bibnamefont{H\"ofer}},
  \bibinfo{author}{\bibfnamefont{I.~L.} \bibnamefont{Shumay}},
  \bibinfo{author}{\bibfnamefont{C.}~\bibnamefont{Reuss}},
  \bibinfo{author}{\bibfnamefont{U.}~\bibnamefont{Thomann}},
  \bibinfo{author}{\bibfnamefont{W.}~\bibnamefont{Wallauer}}, \bibnamefont{and}
  \bibinfo{author}{\bibfnamefont{T.}~\bibnamefont{Fauster}},
  \bibinfo{journal}{Science} \textbf{\bibinfo{volume}{277}},
  \bibinfo{pages}{1480} (\bibinfo{year}{1997}).

\bibitem[{\citenamefont{Hoefer}(1999)}]{Hoefer99}
\bibinfo{author}{\bibfnamefont{U.}~\bibnamefont{H\"ofer}},
  \bibinfo{journal}{Appl. Phys. B} \textbf{\bibinfo{volume}{68}},
  \bibinfo{pages}{383} (\bibinfo{year}{1999}).

\bibitem[{\citenamefont{Lehmann et~al.}(1999)\citenamefont{Lehmann, Merschdorf,
  Thon, Voll, and Pfeiffer}}]{LMT99}
\bibinfo{author}{\bibfnamefont{J.}~\bibnamefont{Lehmann}},
  \bibinfo{author}{\bibfnamefont{M.}~\bibnamefont{Merschdorf}},
  \bibinfo{author}{\bibfnamefont{A.}~\bibnamefont{Thon}},
  \bibinfo{author}{\bibfnamefont{S.}~\bibnamefont{Voll}}, \bibnamefont{and}
  \bibinfo{author}{\bibfnamefont{W.}~\bibnamefont{Pfeiffer}},
  \bibinfo{journal}{Phys. Rev. B} \textbf{\bibinfo{volume}{60}},
  \bibinfo{pages}{17037} (\bibinfo{year}{1999}).

\bibitem[{\citenamefont{Loh et~al.}(1999)\citenamefont{Loh, Sakaguchi, Gamo,
  Tagawa, Sugino, and Ando}}]{LSG99}
\bibinfo{author}{\bibfnamefont{K.~P.} \bibnamefont{Loh}},
  \bibinfo{author}{\bibfnamefont{I.}~\bibnamefont{Sakaguchi}},
  \bibinfo{author}{\bibfnamefont{M.~N.} \bibnamefont{Gamo}},
  \bibinfo{author}{\bibfnamefont{S.}~\bibnamefont{Tagawa}},
  \bibinfo{author}{\bibfnamefont{T.}~\bibnamefont{Sugino}}, \bibnamefont{and}
  \bibinfo{author}{\bibfnamefont{T.}~\bibnamefont{Ando}},
  \bibinfo{journal}{Appl. Phys. Lett.} \textbf{\bibinfo{volume}{74}},
  \bibinfo{pages}{28} (\bibinfo{year}{1999}).

\bibitem[{\citenamefont{Baumeier et~al.}(2007)\citenamefont{Baumeier, Krueger,
  and Pollmann}}]{BKP07}
\bibinfo{author}{\bibfnamefont{B.}~\bibnamefont{Baumeier}},
  \bibinfo{author}{\bibfnamefont{P.}~\bibnamefont{Kruger}}, \bibnamefont{and}
  \bibinfo{author}{\bibfnamefont{J.}~\bibnamefont{Pollmann}},
  \bibinfo{journal}{Phys. Rev. B} \textbf{\bibinfo{volume}{76}},
  \bibinfo{pages}{205404} (\bibinfo{year}{2007}).

\bibitem[{\citenamefont{Emeleus and Coulter}(1987)}]{EC87}
\bibinfo{author}{\bibfnamefont{K.~G.} \bibnamefont{Emeleus}} \bibnamefont{and}
  \bibinfo{author}{\bibfnamefont{J.~R.~M.} \bibnamefont{Coulter}},
  \bibinfo{journal}{Int. J. Electronics} \textbf{\bibinfo{volume}{62}},
  \bibinfo{pages}{225} (\bibinfo{year}{1987}).

\bibitem[{\citenamefont{Emeleus and Coulter}(1988)}]{EC88}
\bibinfo{author}{\bibfnamefont{K.~G.} \bibnamefont{Emeleus}} \bibnamefont{and}
  \bibinfo{author}{\bibfnamefont{J.~R.~M.} \bibnamefont{Coulter}},
  \bibinfo{journal}{IEE Proceedings} \textbf{\bibinfo{volume}{135}},
  \bibinfo{pages}{76} (\bibinfo{year}{1988}).

\bibitem[{\citenamefont{Behnke et~al.}(1997)\citenamefont{Behnke, Bindemann,
  Deutsch, and Becker}}]{BBD97}
\bibinfo{author}{\bibfnamefont{J.~F.} \bibnamefont{Behnke}},
  \bibinfo{author}{\bibfnamefont{T.}~\bibnamefont{Bindemann}},
  \bibinfo{author}{\bibfnamefont{H.}~\bibnamefont{Deutsch}}, \bibnamefont{and}
  \bibinfo{author}{\bibfnamefont{K.}~\bibnamefont{Becker}},
  \bibinfo{journal}{Contrib. Plasma Phys.} \textbf{\bibinfo{volume}{37}},
  \bibinfo{pages}{345} (\bibinfo{year}{1997}).

\bibitem[{\citenamefont{Golubovskii et~al.}(2002)\citenamefont{Golubovskii,
  Maiorov, Behnke, and Behnke}}]{GMB02}
\bibinfo{author}{\bibfnamefont{Y.~B.} \bibnamefont{Golubovskii}},
  \bibinfo{author}{\bibfnamefont{V.~A.} \bibnamefont{Maiorov}},
  \bibinfo{author}{\bibfnamefont{J.}~\bibnamefont{Behnke}}, \bibnamefont{and}
  \bibinfo{author}{\bibfnamefont{J.~F.} \bibnamefont{Behnke}},
  \bibinfo{journal}{J. Phys. D: Appl. Phys} \textbf{\bibinfo{volume}{35}},
  \bibinfo{pages}{751} (\bibinfo{year}{2002}).

\bibitem[{\citenamefont{Bronold et~al.}(2008)\citenamefont{Bronold, Fehske,
  Kersten, and Deutsch}}]{BFKD08}
\bibinfo{author}{\bibfnamefont{F.~X.} \bibnamefont{Bronold}},
  \bibinfo{author}{\bibfnamefont{H.}~\bibnamefont{Fehske}},
  \bibinfo{author}{\bibfnamefont{H.}~\bibnamefont{Kersten}}, \bibnamefont{and}
  \bibinfo{author}{\bibfnamefont{H.}~\bibnamefont{Deutsch}},
  \bibinfo{journal}{Phys. Rev. Lett.} \textbf{\bibinfo{volume}{101}},
  \bibinfo{pages}{175002} (\bibinfo{year}{2008}).

\bibitem[{\citenamefont{Bronold et~al.}(2009)\citenamefont{Bronold, Deutsch,
  and Fehske}}]{BDF09}
\bibinfo{author}{\bibfnamefont{F.~X.} \bibnamefont{Bronold}},
  \bibinfo{author}{\bibfnamefont{H.}~\bibnamefont{Deutsch}}, \bibnamefont{and}
  \bibinfo{author}{\bibfnamefont{H.}~\bibnamefont{Fehske}},
  \bibinfo{journal}{Eur. Phys. J. D} \textbf{\bibinfo{volume}{54}},
  \bibinfo{pages}{519} (\bibinfo{year}{2009}).

\bibitem[{\citenamefont{Bronold et~al.}(2011)\citenamefont{Bronold, Heinisch,
  Marbach, and Fehske}}]{BHMF10}
\bibinfo{author}{\bibfnamefont{F.~X.} \bibnamefont{Bronold}},
  \bibinfo{author}{\bibfnamefont{R.~L.} \bibnamefont{Heinisch}},
  \bibinfo{author}{\bibfnamefont{J.}~\bibnamefont{Marbach}}, \bibnamefont{and}
  \bibinfo{author}{\bibfnamefont{H.}~\bibnamefont{Fehske}},
  \bibinfo{journal}{IEEE Transactions on Plasma Science}
  \textbf{\bibinfo{volume}{xx}}, \bibinfo{pages}{xx} (\bibinfo{year}{2011}).

\bibitem[{\citenamefont{Geis et~al.}(2005)\citenamefont{Geis, Deneault, Krohn,
  Marchant, Lyszczarz, and Cooke}}]{GDK05}
\bibinfo{author}{\bibfnamefont{M.~W.} \bibnamefont{Geis}},
  \bibinfo{author}{\bibfnamefont{S.}~\bibnamefont{Deneault}},
  \bibinfo{author}{\bibfnamefont{K.~E.} \bibnamefont{Krohn}},
  \bibinfo{author}{\bibfnamefont{M.}~\bibnamefont{Marchant}},
  \bibinfo{author}{\bibfnamefont{T.~M.} \bibnamefont{Lyszczarz}},
  \bibnamefont{and} \bibinfo{author}{\bibfnamefont{D.~L.} \bibnamefont{Cooke}},
  \bibinfo{journal}{Appl. Phys. Lett.} \textbf{\bibinfo{volume}{87}},
  \bibinfo{pages}{192115} (\bibinfo{year}{2005}).

\bibitem[{\citenamefont{Biasini et~al.}(2005)\citenamefont{Biasini, Gann,
  Yarmoff, Mills, Pfeiffer, West, Gao, and Williams}}]{BGY05}
\bibinfo{author}{\bibfnamefont{M.}~\bibnamefont{Biasini}},
  \bibinfo{author}{\bibfnamefont{R.~D.} \bibnamefont{Gann}},
  \bibinfo{author}{\bibfnamefont{J.~A.} \bibnamefont{Yarmoff}},
  \bibinfo{author}{\bibfnamefont{A.~P.} \bibnamefont{Mills}},
  \bibinfo{author}{\bibfnamefont{L.~N.} \bibnamefont{Pfeiffer}},
  \bibinfo{author}{\bibfnamefont{K.~W.} \bibnamefont{West}},
  \bibinfo{author}{\bibfnamefont{X.~P.~W.} \bibnamefont{Gao}},
  \bibnamefont{and} \bibinfo{author}{\bibfnamefont{B.~C.~D.}
  \bibnamefont{Williams}}, \bibinfo{journal}{Appl. Phys. Lett.}
  \textbf{\bibinfo{volume}{86}}, \bibinfo{pages}{162111}
  (\bibinfo{year}{2005}).

\bibitem[{\citenamefont{Hattori}(1995)}]{Hattori95}
\bibinfo{author}{\bibfnamefont{H.}~\bibnamefont{Hattori}},
  \bibinfo{journal}{Chem. Rev.} \textbf{\bibinfo{volume}{95}},
  \bibinfo{pages}{537} (\bibinfo{year}{1995}).

\bibitem[{\citenamefont{Hattori}(2003)}]{Hattori03}
\bibinfo{author}{\bibfnamefont{H.}~\bibnamefont{Hattori}}, \bibinfo{journal}{J.
  Jpn. Petrol. Inst.} \textbf{\bibinfo{volume}{47}}, \bibinfo{pages}{67}
  (\bibinfo{year}{2003}).

\bibitem[{\citenamefont{Wolf and Ertl}(2000)}]{WE00}
\bibinfo{author}{\bibfnamefont{M.}~\bibnamefont{Wolf}} \bibnamefont{and}
  \bibinfo{author}{\bibfnamefont{G.}~\bibnamefont{Ertl}},
  \bibinfo{journal}{Science} \textbf{\bibinfo{volume}{288}},
  \bibinfo{pages}{1352} (\bibinfo{year}{2000}).

\bibitem[{\citenamefont{Freund}(2007)}]{Freund07}
\bibinfo{author}{\bibfnamefont{H.-J.} \bibnamefont{Freund}},
  \bibinfo{journal}{Surface science} \textbf{\bibinfo{volume}{601}},
  \bibinfo{pages}{1438} (\bibinfo{year}{2007}).

\bibitem[{\citenamefont{Heinisch
  et~al.}(2010{\natexlab{a}})\citenamefont{Heinisch, Bronold, and
  Fehske}}]{HBF10a}
\bibinfo{author}{\bibfnamefont{R.~L.} \bibnamefont{Heinisch}},
  \bibinfo{author}{\bibfnamefont{F.~X.} \bibnamefont{Bronold}},
  \bibnamefont{and} \bibinfo{author}{\bibfnamefont{H.}~\bibnamefont{Fehske}},
  \bibinfo{journal}{Phys. Rev. B} \textbf{\bibinfo{volume}{81}},
  \bibinfo{pages}{155420} (\bibinfo{year}{2010}{\natexlab{a}}).

\bibitem[{\citenamefont{Heinisch
  et~al.}(2010{\natexlab{b}})\citenamefont{Heinisch, Bronold, and
  Fehske}}]{HBF10b}
\bibinfo{author}{\bibfnamefont{R.~L.} \bibnamefont{Heinisch}},
  \bibinfo{author}{\bibfnamefont{F.~X.} \bibnamefont{Bronold}},
  \bibnamefont{and} \bibinfo{author}{\bibfnamefont{H.}~\bibnamefont{Fehske}},
  \bibinfo{journal}{Phys. Rev. B} \textbf{\bibinfo{volume}{82}},
  \bibinfo{pages}{125408} (\bibinfo{year}{2010}{\natexlab{b}}).

\bibitem[{\citenamefont{Bendow and Ying}(1973)}]{BY73}
\bibinfo{author}{\bibfnamefont{B.}~\bibnamefont{Bendow}} \bibnamefont{and}
  \bibinfo{author}{\bibfnamefont{S.-C.} \bibnamefont{Ying}},
  \bibinfo{journal}{Phys. Rev. B} \textbf{\bibinfo{volume}{7}},
  \bibinfo{pages}{622} (\bibinfo{year}{1973}).

\bibitem[{\citenamefont{Gortel et~al.}(1980{\natexlab{a}})\citenamefont{Gortel,
  Kreuzer, and Teshima}}]{GKT80b}
\bibinfo{author}{\bibfnamefont{Z.~W.} \bibnamefont{Gortel}},
  \bibinfo{author}{\bibfnamefont{H.~J.} \bibnamefont{Kreuzer}},
  \bibnamefont{and} \bibinfo{author}{\bibfnamefont{R.}~\bibnamefont{Teshima}},
  \bibinfo{journal}{Phys. Rev. B} \textbf{\bibinfo{volume}{22}},
  \bibinfo{pages}{512} (\bibinfo{year}{1980}{\natexlab{a}}).

\bibitem[{\citenamefont{Armand and Manson}(1991)}]{AM91}
\bibinfo{author}{\bibfnamefont{G.}~\bibnamefont{Armand}} \bibnamefont{and}
  \bibinfo{author}{\bibfnamefont{J.~R.} \bibnamefont{Manson}},
  \bibinfo{journal}{Phys. Rev. B} \textbf{\bibinfo{volume}{43}},
  \bibinfo{pages}{14371} (\bibinfo{year}{1991}).

\bibitem[{\citenamefont{Gortel et~al.}(1980{\natexlab{b}})\citenamefont{Gortel,
  Kreuzer, and Teshima}}]{GKT80a}
\bibinfo{author}{\bibfnamefont{Z.~W.} \bibnamefont{Gortel}},
  \bibinfo{author}{\bibfnamefont{H.~J.} \bibnamefont{Kreuzer}},
  \bibnamefont{and} \bibinfo{author}{\bibfnamefont{R.}~\bibnamefont{Teshima}},
  \bibinfo{journal}{Phys. Rev. B} \textbf{\bibinfo{volume}{22}},
  \bibinfo{pages}{5655} (\bibinfo{year}{1980}{\natexlab{b}}).

\bibitem[{\citenamefont{Kreuzer and Gortel}(1986)}]{KG86}
\bibinfo{author}{\bibfnamefont{H.~J.} \bibnamefont{Kreuzer}} \bibnamefont{and}
  \bibinfo{author}{\bibfnamefont{Z.~W.} \bibnamefont{Gortel}},
  \emph{\bibinfo{title}{Physisorption Kinetics}} (\bibinfo{publisher}{Springer
  Verlag}, \bibinfo{address}{Berlin}, \bibinfo{year}{1986}).

\bibitem[{\citenamefont{Iche and Nozi\'{e}res}(1976)}]{IN76}
\bibinfo{author}{\bibfnamefont{G.}~\bibnamefont{Iche}} \bibnamefont{and}
  \bibinfo{author}{\bibfnamefont{P.}~\bibnamefont{Nozi\'{e}res}},
  \bibinfo{journal}{J. Phys. (Paris)} \textbf{\bibinfo{volume}{37}},
  \bibinfo{pages}{1313} (\bibinfo{year}{1976}).

\bibitem[{\citenamefont{Brenig}(1982)}]{Brenig82}
\bibinfo{author}{\bibfnamefont{W.}~\bibnamefont{Brenig}}, \bibinfo{journal}{Z.
  Phys. B} \textbf{\bibinfo{volume}{48}}, \bibinfo{pages}{127}
  (\bibinfo{year}{1982}).

\bibitem[{\citenamefont{Evans and Mills}(1973)}]{EM73}
\bibinfo{author}{\bibfnamefont{E.}~\bibnamefont{Evans}} \bibnamefont{and}
  \bibinfo{author}{\bibfnamefont{D.~L.} \bibnamefont{Mills}},
  \bibinfo{journal}{Phys. Rev. B} \textbf{\bibinfo{volume}{8}},
  \bibinfo{pages}{4004} (\bibinfo{year}{1973}).

\bibitem[{\citenamefont{\v{S}iber and Gumhalter}(2003)}]{SG03}
\bibinfo{author}{\bibfnamefont{A.}~\bibnamefont{\v{S}iber}} \bibnamefont{and}
  \bibinfo{author}{\bibfnamefont{B.}~\bibnamefont{Gumhalter}},
  \bibinfo{journal}{Phys. Rev. Lett.} \textbf{\bibinfo{volume}{90}},
  \bibinfo{pages}{126103} (\bibinfo{year}{2003}).

\bibitem[{\citenamefont{\v{S}iber and Gumhalter}(2005)}]{SG05}
\bibinfo{author}{\bibfnamefont{A.}~\bibnamefont{\v{S}iber}} \bibnamefont{and}
  \bibinfo{author}{\bibfnamefont{B.}~\bibnamefont{Gumhalter}},
  \bibinfo{journal}{Phys. Rev. B} \textbf{\bibinfo{volume}{71}},
  \bibinfo{pages}{081401} (\bibinfo{year}{2005}).

\bibitem[{\citenamefont{\v{S}iber and Gumhalter}(2008)}]{SG08}
\bibinfo{author}{\bibfnamefont{A.}~\bibnamefont{\v{S}iber}} \bibnamefont{and}
  \bibinfo{author}{\bibfnamefont{B.}~\bibnamefont{Gumhalter}},
  \bibinfo{journal}{J. Phys. Condens. Matter} \textbf{\bibinfo{volume}{20}},
  \bibinfo{pages}{224002} (\bibinfo{year}{2008}).

\bibitem[{\citenamefont{Manson}(1991)}]{Manson91}
\bibinfo{author}{\bibfnamefont{J.~R.} \bibnamefont{Manson}},
  \bibinfo{journal}{Phys. Rev. B} \textbf{\bibinfo{volume}{43}},
  \bibinfo{pages}{6924} (\bibinfo{year}{1991}).

\bibitem[{\citenamefont{Glauber}(1955)}]{Glauber55}
\bibinfo{author}{\bibfnamefont{R.~J.} \bibnamefont{Glauber}},
  \bibinfo{journal}{Phys. Rev.} \textbf{\bibinfo{volume}{98}},
  \bibinfo{pages}{1692} (\bibinfo{year}{1955}).

\end{thebibliography}
%\bibliographystyle{apsrev}

\end{document}